\documentclass{article}[12pt]
\usepackage{setspace}
\usepackage[margin=1in]{geometry}
\usepackage{booktabs}
\usepackage{color}
\usepackage{graphicx}
\usepackage{amssymb}
\usepackage{caption}
\usepackage{threeparttable}
\usepackage{authblk}
\usepackage{hyperref}

\usepackage{natbib}

\begin{document}\sloppy

\author[1]{Leslie Myint}
\author[2]{Aboozar Hadavand}
\author[2]{Leah Jager}
\author[2]{Jeffrey Leek}

\affil[1]{Department of Mathematics, Statistics, and Computer Science, Macalester College, 1600 Grand Ave, Saint Paul, MN 55105}
\affil[2]{Department of Biostatistics, Johns Hopkins Bloomberg School of Public Health, 615 N. Wolfe St, Baltimore, MD 21212}

\title{Comparison of plotting system outputs in beginner analysts}
\date{}
\maketitle

\begin{abstract}
The R programming language is built on an ecosystem of packages, some that allow analysts to accomplish the same tasks. For example, there are at least two clear workflows for creating data visualizations in R: using the base graphics package (referred to as ``base R'') and the ggplot2 add-on package based on the grammar of graphics. Here we perform an empirical study of the quality of scientific graphics produced by beginning R users. In our experiment, learners taking a data science course on the Coursera platform were randomized to complete identical plotting exercises in either the base R or the ggplot2 system. Learners were then asked to evaluate their peers in terms of visual characteristics key to scientific cognition. We observed that graphics created with the two systems rated similarly on many characteristics. However, ggplot2 graphics were generally judged to be more visually pleasing and, in the case of faceted scientific plots, easier to understand. Our results suggest that while both graphic systems are useful in the hands of beginning users, ggplot2's natural faceting system may be easier to use by beginning users for displaying more complex relationships.
\end{abstract}



\section{Introduction}\label{intro}

The R programming language is one of the most popular means of introducing computing into data science, data analytics, and statistics curricula \citep{Cetinkaya-Rundel2018}. An advantage of the R ecosystem is the powerful set of add-on packages that can be used to perform a range of tasks from experimental design \citep{doe_cran_taskview}, to data cleaning \citep{tidy_peerj, r4ds}, to visualization \citep{ggplot2_book}, and modeling or machine learning \citep{caret}. 

While these packages make it possible for users of R to accomplish a wide range of tasks, it also means there are often multiple workflows for accomplish the same data analytic goal. These competing workflows often lead to strong opinions and debates in the literature, on social media, and on blogs \citep{leek_ss,drob_why_gg}. But we have collected relatively little information about the way that these tools are used in the hands of end users. 

One of the more commonly debated aspects of data science education within the R community is the plotting system used to introduce learners to statistical graphics. Generally, the two main systems under consideration are the base graphics package in R (called ``base R'') and the ggplot2 graphics systems based on the grammar of graphics \citep{ggplot2_book}. There has been some online and informal debate about the general strengths and weaknesses of these two systems for both research and teaching \citep{leek_ss,drob_why_gg}. More recently there has been discussion of the relative merits of the two plotting systems in teaching the specific student population of beginner analysts \citep{drob_teach_gg} and some investigation of learning outcomes when using base R and ggplot2 in the classroom \citep{Stander2017}. In the latter investigation, Stander et al. provide instruction in both plotting systems in the classroom but do not formally compare the systems in terms of student learning outcomes.

There has also been a surge in the creation of resources that focus on ggplot2, and more broadly, the encompassing tidyverse framework \citep{r4ds}. The Modern Dive open source introductory textbook for data science education with R is one such example \citep{moderndive}. \citet{tidy_peerj} describe a full data analytic workflow in this framework. Generally, proponents of the ggplot2 system cite the harmonization of the tidy data mindset and ggplot2 syntax for mapping between variables and visual plot elements. They also appeal to the modular nature of the syntax that gives rise to the ability to build plots in layers. Proponents of the base R system cite its power to create nearly any imaginable graphic by acting on individual plot elements such as points and lines. Making plots in the base R system also increases exposure to for-loops (and related ideas), which can be helpful to students in other aspects of their data science training.

A primary goal in statistical education is giving students the ability to communicate effectively with data. Of course, statistical graphics are a major part of effective data communication, and a large body of literature on the visual display of scientific research and human cognition highlights the need for thinking critically about statistical graphics education. For one, 
\citet{tversky2002animation} argue that a visual display must be accurately perceived to be effective and refer to this as the ``apprehension principle of visual displays''. In the context of presentation of physical processes, they show that animations are not more effective than static graphics. \citet{smallman2005naive} provide a similar analysis in the context of visual dimensions. They find that people misperceive distances in depth and, therefore, 3D displays are not ideal for presenting absolute distances. \citet{kosslyn2006graph} argues that graphics should not present information beyond what is needed by the user. \citet{rosenholtz2007measuring} and \citet{wickens1995proximity} find that presenting too much information in the display can lead to visual distraction in non-expert audiences of scientific research. Therefore, it is important for statistical educators to teach graphics systems that aid students' creation of effective data displays - that is, data displays that enhance scientific cognition.

Here we seek to better understand differences in the visual display and perception of plots made in the base R and ggplot2 systems. We study this in a group of beginner learners within the Coursera platform. Specifically, we report results from a randomized experiment in which learners were randomized to complete identical plotting exercises in either the base R or the ggplot2 system. Learners were then asked to evaluate plots from their peers in terms of visual characteristics key to scientific cognition.

We hypothesized that plots made with ggplot2 would generally rate higher on aesthetics and clarity due to the relative ease of the syntax and the default layout. That is, we believed that it would be syntactically easier for students to create ``correct'' or effective plots in ggplot2. At the same time, we hypothesized that plots in base R would show clearer labels due to its undesirable default labels. We suspected that students' direct modification of labels (as opposed to accepting defaults) would result in higher clarity labeling.

We find that, for the specific exercises given to the students, the aesthetic differences between the two plotting systems (as measured by the peer review) are generally small. However, we find that the plots made with ggplot2 are generally of higher clarity than those made in the base R system, particularly when the students were asked to make a complex, multi-panel plot. We also observe differences between the systems in the number of panels used in this complex, multi-plot, suggesting different cognitive interactions with the R syntax.

\section{Methods}\label{methods}

We ran a randomized experiment from July 2016 to September 2017 within the Reproducible Research course in the Johns Hopkins Data Science Specialization on Coursera. This course covers the basics of RMarkdown, literate programming, and the principles of reproducible research. This course follows Exploratory Data Analysis, a course that covers the base R and ggplot2 systems as well as concepts involved in thorough exploratory analysis. Since the launch of Reproducible Research, 187,617 learners have enrolled, from which 29,534 have completed the course. Demographic information summaries are available in Table~\ref{tab:demog}. This demographic information is specific to this offering of the Reproducible Research course on Coursera, but but not necessarily specific to the students who participated in the experiment.

\begin{table}
\centering
  \begin{threeparttable}
\footnotesize
\caption{Learner demographics in the Reproducible Research course}\
\centering
\label{tab:demog}
\begin{tabular}{p{4cm}|p{6cm}}
Characteristics & Shares \\
\addlinespace
\toprule
Gender & Male: 76\%\\
       & Female: 24\%\\
\midrule
Student status & Non-degree student: 68\%\\
           & Full-time student: 24\%\\
           & Part-time student: 8\%\\
\midrule
Education & College (no degree): 4\%\\
           & Bachelor's degree: 34\%\\
           & Master's degree: 46\%\\
           & Doctorate degree: 11\%\\
           & Other: 5\%\\
\midrule
Employment Status & Full-time: 68\%\\
           & Part-time: 4\%\\
           & Unemployed (looking for work): 16\%\\
           & Other: 12\%\\
\midrule
Language & English: 89\%\\
           & Chinese: 3\%\\
           & Other: 8\%\\
\midrule
Country & United States: 36\%\\
           & India: 12\%\\
           & Great Britain: 4\%\\
           & Canada: 3\%\\
           & Germany: 3\%\\
           & China: 3\%\\
           & Other: 39\%\\
\addlinespace
\bottomrule
\end{tabular}
\begin{tablenotes}
      \scriptsize
      \item Note: The demographic information is for all students who took the course Reproducible Research as part of the Johns Hopkins Data Science Specialization on Coursera. It is not necessarily specific to the students who took participated in our experiment.
    \end{tablenotes}
\end{threeparttable}
\end{table}

The Coursera platform allowed us to randomize two versions of a peer-graded assignment across learners. All students had the option of completing a peer-graded assignment involving the creation of two plots: one showing the relationship between two continuous variables and one showing how this relationship varied across strata of two categorical variables. The data given to the students contained information on medical charges and insurance payments, which were the two continuous variables. The data also contained information on 6 states and 6 medical conditions. These were the two categorical variables that students were to use in the second plot. In total there are 36 state-medical condition combinations. The first plot will be referred to as the ``simple'' plot, and the second will be referred to as the ``complex'' plot. The assignment is shown below for the base R arm:

\begin{quote}
\textit{To practice the plotting techniques you have learned so far, you will be making a graphic that explores relationships between variables. You will be looking at a subset of a United States medical expenditures dataset with information on costs for different medical conditions and in different areas of the country.}

\textit{You should do the following:}
\begin{enumerate}
\item \textit{Make a plot that answers the question: what is the relationship between mean covered charges (\texttt{Average.Covered.Charges}) and mean total payments (\texttt{Average.Total.Payments}) in New York?}
\item \textit{Make a plot (possibly multi-panel) that answers the question: how does the relationship between mean covered charges (\texttt{Average.Covered.Charges}) and mean total payments (\texttt{Average.Total.Payments}) vary by medical condition (\texttt{DRG.Definition}) and the state in which care was received (\texttt{Provider.State})?}
\end{enumerate}

\textbf{\textit{Use only the base graphics system to make your figure.}}
\textit{Please submit to the peer assessment two PDF files, one for each of the two plots. You will be graded on whether you answered the questions and a number of features describing the clarity of the plots including axis labels, figure legends, figure captions, and plots. For guidelines on how to create production quality plots see Chapter 10 of the Elements of Data Analytic Style.}
\end{quote}

\begin{figure}
  \centering
  \captionsetup{width=.9\linewidth}
  \includegraphics[width=\textwidth]{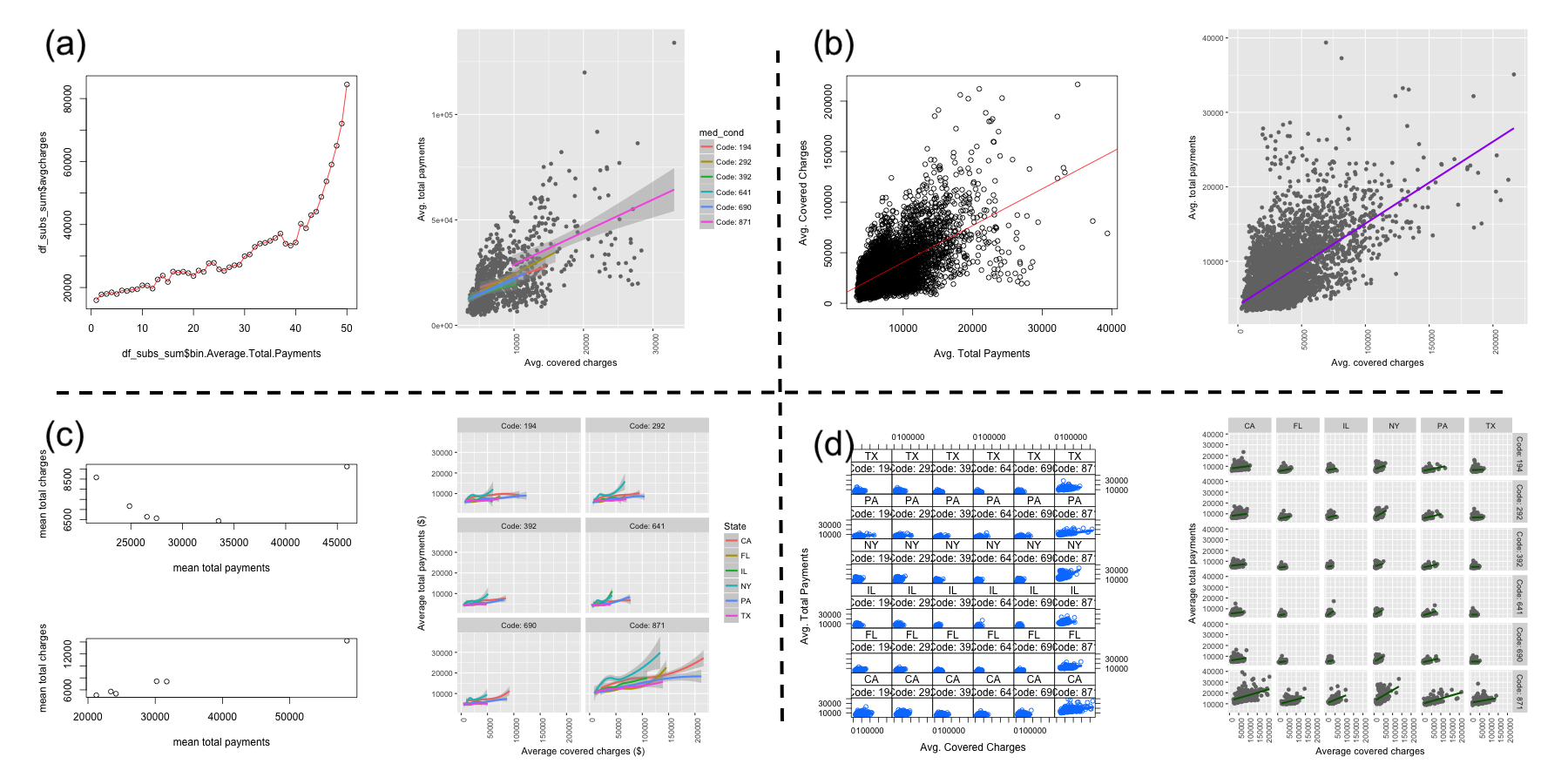}
  \caption{\textbf{Sample student submissions.} The top panels show example student submissions for the simple plot that are of low (a) and high (b) quality. The bottom panels show example submissions for the complex plot that are of low (c) and high (d) quality. The left hand plots in each section were made in base R, and the right hand plots were made in ggplot2. For privacy reasons, none of these figures were actually made by students. These figures are recreations that show general types of figures that were commonly made by students.}
  \label{fig:samplesubmission}
\end{figure}

In the ggplot2 arm of the experiment, learners instead saw the sentence: \textbf{Use only the ggplot2 graphics system to make your figure.} Figure \ref{fig:samplesubmission} shows sample submissions for both arms of the study and for both the simple and complex questions. We emphasize that as part of the assignment prompt, we tell students how their plots will be evaluated. We tell them that they will be evaluated on answering the questions, including axis labels, figure legends, figure captions, and on the type of plot created. We also point them to a resource from a previous course in the specialization that discusses exactly these points. That all students are aware of this before submitting means that students in both arms had opportunities to specifically work on these aspects of their plots. Thus differences between arms are not attributable to lack of awareness about assessment criteria but more so to students' skill with the two plotting systems.

After completing the assignment, students were asked to review one or more assignments from their peers. Students reviewed assignments that used the same plotting system in which they completed their assignment. The review rubric is shown in Table~\ref{tab:peer_review}. For the question, ``Did they upload a plot?'', we provide three response choices for the simple plot to determine if the student uploaded a plot using the correct plotting system. Because the peer review rubric starts with assessments for the simple plot, we operated under the assumption that this compliance status also applied to the complex plot. For this reason, there are only two answer choices for this question for the complex plot.

The complex plot involved visualizing a relationship between two continuous variables in the 36 (states/medical conditions) subgroups. For the complex plots only, we manually annotated certain visual features of the uploaded plots. These annotations included 3 pieces of information: 

\begin{enumerate}
    \item Our own judgment of whether the plot was made with the correct plotting system. (Correct/Incorrect)
    \item The number of panels present in the plot (corresponding to subgroups). Most common values were 1, 2, 6, 12, and 36.
    \item An indication (yes/no) of whether the plot had some other visual grouping that was not a panel. For example, points within one panel could be colored by medical condition to satisfy this criterion.
\end{enumerate}

\begin{table}[]
\centering
\footnotesize
\caption{Peer review rubric}
\label{tab:peer_review}
\begin{tabular}[t]{l|ll}
\toprule
\textbf{Plot aspect} & \multicolumn{2}{l}{\textbf{Question/Answer Choices}} \\
\, & \textbf{Plot 1} & \textbf{Plot 2} \\
\hline
Presence & \begin{tabular}[t]{@{}l@{}}Did they upload a plot?\\ No\\ Yes\\ Yes, and they made it \\ with (base R)(ggplot2).\end{tabular} & \begin{tabular}[t]{@{}l@{}}Did they upload a plot?\\ No\\ Yes\end{tabular} \\
\hline
Content & \begin{tabular}[t]{@{}l@{}}Does the plot clearly show the \\ relationship between mean covered \\ charges (Average.Covered.Charges) \\ and mean total payments \\ (Average.Total.Payments) in\\ New York?\\ No\\ Yes\end{tabular} & \begin{tabular}[t]{@{}l@{}}Does the plot clearly show the \\ relationship between mean covered \\ charges (Average.Covered.Charges) \\ and mean total payments \\ (Average.Total.Payments) vary by \\ medical condition (DRG.Definition) \\ and the state in which care was \\ received (Provider.State)?\\ No\\ Yes\end{tabular} \\
\hline
General aesthetics & \multicolumn{2}{l}{\begin{tabular}[t]{@{}l@{}}Is the plot visually pleasing?\\ No\\ Yes\end{tabular}} \\
\hline
Clarity & \multicolumn{2}{l}{\begin{tabular}[t]{@{}l@{}}Can the plot be understood without a figure caption?\\ No\\ Yes\end{tabular}} \\
\hline
Annotation & \multicolumn{2}{l}{\begin{tabular}[t]{@{}l@{}}Are the legends and labels sufficient to explain what the plot is showing?\\ No\\ Yes\end{tabular}} \\
\hline
Display & \multicolumn{2}{l}{\begin{tabular}[t]{@{}l@{}}Are the plot text and labels large enough to read?\\ No\\ Yes\end{tabular}} \\
\hline
Annotation & \multicolumn{2}{l}{\begin{tabular}[t]{@{}l@{}}Do the plot text and labels use full words instead of abbreviations?\\ No\\ Yes\end{tabular}} \\
\bottomrule
\end{tabular}
\end{table}

\section{Results}\label{results}

A total of 1078 students participated in the trial. In the base R arm, 436 students submitted a plot, and in the ggplot2 arm, 642 students participated. This differential participation could be due to students feeling less comfortable with base R graphics than ggplot2, but in the absence of information on characteristics and course outcomes for these non-participants, we analyzed data for these 1078 students who completed the assignment and peer review. Among these students, there was a 100 percent response rate for all items on the review rubric. Each student was asked to review the submissions of at least one and possibly multiple other students. There were 1267 total peer reviews in the base R arm and 1440 in the ggplot2 arm. In the following results, we remove peer review responses for which the reviewer answered ``No'' to the ``Did they upload a plot?'' question.

\subsection{Compliance}

Based on our manual annotation of the complex plots in both the base R and ggplot2 arms, we were able to compute the exact percentage of complex plot submissions that were made in the correct plotting system. In the base R arm, 433 of the 436 submitted plots could be annotated. (The remaining 3 plots were completely empty files.) Of the 433 annotated plots, 375 (86.6\%) were made in the base R system. In the ggplot2 arm, 637 of the 642 submitted plots could be annotated. (The remaining 5 plots were completely empty files.) Of the 637 annotated plots, 636 (99.8\%) were made in the ggplot2 system. The higher rate of compliance for the complex plot in the ggplot2 arm was expected given the more concise syntax of the ggplot2 system.

We did not annotate the simple plots but we expect that rates of compliance would be similar to that for the complex plots. We are still able to estimate compliance rates for the simple plot through the peer review question ``Did they upload a plot?'' Student reviewers were able to choose from ``No'', ``Yes'', and ``Yes, and they made it with base R (ggplot2).'' In each arm, we estimate the compliance rate to be the fraction of the time the third response was chosen in all peer reviews. We estimate the compliance rate for the simple plot to be 92.9\% in the base R arm and 97.3\% in the ggplot2 arm.

\subsection{Visual characteristics of submitted plots}

Peer review outcomes for all students are displayed in Table~\ref{tab:review_responses_comparison}. Review outcomes for visual characteristics were similar between the base R and ggplot2 systems. For most characteristics, the systems differed by only a few percentage points, but positive plot qualities were more likely to be seen in plots made in ggplot2. Further, positive qualities were more likely to be seen in the simple plot than in the complex plot for both systems.

In terms of general aesthetics, plots made in ggplot2 were more likely to be viewed as visually pleasing, and this difference was more pronounced in the simple plot in the complex plot. 

Ratings of overall clarity (``Does the plot clearly show the relationship?'') were higher for figures made in ggplot2 for both the simple and complex plots, and the difference between the systems was larger for the complex plot. We also assessed plot clarity through the two questions: ``Can the plot be understood without a figure caption?'' and ``Are the legends and labels sufficient to explain what the plot is showing?''. For these two questions, the difference between the two systems is more pronounced for the complex plot. For the complex plot, submissions made in ggplot2 were more likely to be perceived as being sufficiently clear as a standalone figure.

For both the simple and complex plots, there is no indication of differences in tendencies to use full words versus abbreviations between the two plotting systems. This is sensible given that users create the text of plot annotations in nearly the same way in both systems. Interestingly, for the complex plot, graphics made in ggplot2 were less likely to have plot text and labels that were large enough to read. This may be due to the nature of text resizing when plotting with facets in ggplot2 and to a lack of instructional time spent on fine tuning such visual aspects within the course.

We also examined peer review outcomes on the subset of students that complied with their assigned plotting system, and we find that results are almost identical to the results discussed above for the full set of reviews (Table~\ref{tab:review_responses_comparison_compliers}).


\begin{table}[ht]
\centering
\begin{threeparttable}
\scriptsize
\caption{\textbf{Comparison of peer review responses in the base R and ggplot2 arms (all student submissions).}}
\label{tab:review_responses_comparison}
\begin{tabular}{lllrrl}
  \toprule
Plot & Prompt & Response & Base R & ggplot2 & ggplot2 - base R \\ 
  \hline
simple & Clearly shows relationship? & Yes & 86.2\% & 89.7\% & 3.5\% (1\%, 6.1\%)* \\ 
  simple & Is the plot visually pleasing? & Somewhat & 23.1\% & 18.3\% & -4.8\% (-8\%, -1.7\%)* \\ 
  simple & Is the plot visually pleasing? & Yes & 73.7\% & 80.5\% & 6.9\% (3.6\%, 10.1\%)* \\ 
  simple & Understandable without caption? & Yes & 90.9\% & 91\% & 0.1\% (-2.1\%, 2.4\%) \\ 
  simple & Legends and labels sufficient? & Yes & 89.4\% & 90.8\% & 1.4\% (-1\%, 3.7\%) \\ 
  simple & Text and labels large enough? & Yes & 97.8\% & 99\% & 1.2\% (0.2\%, 2.2\%)* \\ 
  simple & Use full words vs. abbreviations? & Yes & 95.4\% & 96.1\% & 0.7\% (-0.9\%, 2.3\%) \\
  \hline
  complex & Clearly shows relationships? & Yes & 72.3\% & 83.6\% & 11.4\% (8.1\%, 14.6\%)* \\ 
  complex & Is the plot visually pleasing? & Somewhat & 30\% & 30.8\% & 0.8\% (-2.8\%, 4.3\%) \\ 
  complex & Is the plot visually pleasing? & Yes & 59.5\% & 60.6\% & 1\% (-2.8\%, 4.8\%) \\ 
  complex & Understandable without caption? & Yes & 76.8\% & 81.5\% & 4.7\% (1.5\%, 7.9\%)* \\ 
  complex & Legends and labels sufficient? & Yes & 77.9\% & 82.4\% & 4.5\% (1.4\%, 7.6\%)* \\ 
  complex & Text and labels large enough? & Yes & 89.8\% & 86.3\% & -3.5\% (-6\%, -1\%)* \\ 
  complex & Use full words vs. abbreviations? & Yes & 83.6\% & 85.4\% & 1.8\% (-1\%, 4.6\%) \\ 
   \bottomrule
\end{tabular}
\begin{tablenotes}
      \scriptsize
      \item Note: For each rubric item and response, the percentage of reviews indicating that response are shown. The last column gives the difference between the ggplot2 and base R arms and the 95\% confidence interval for that difference.
    \end{tablenotes}
\end{threeparttable}
\end{table}

\begin{table}[ht]
\centering
\begin{threeparttable}
\scriptsize
\caption{\textbf{Comparison of peer review responses in the base R and ggplot2 arms (compliant submissions).}}
\label{tab:review_responses_comparison_compliers}
\begin{tabular}{lllrrl}
  \toprule
Plot & Prompt & Response & Base R & ggplot2 & ggplot2 - base R \\ 
  \hline
simple & Clearly shows relationship? & Yes & 85.7\% & 89.7\% & 4\% (1.3\%, 6.7\%)* \\ 
  simple & Is the plot visually pleasing? & Somewhat & 23.1\% & 18.3\% & -4.8\% (-8.1\%, -1.5\%)* \\ 
  simple & Is the plot visually pleasing? & Yes & 73.6\% & 80.5\% & 6.9\% (3.5\%, 10.3\%)* \\ 
  simple & Understandable without caption? & Yes & 90.6\% & 90.9\% & 0.3\% (-2.1\%, 2.6\%) \\ 
  simple & Legends and labels sufficient? & Yes & 89.3\% & 90.7\% & 1.3\% (-1.1\%, 3.8\%) \\ 
  simple & Text and labels large enough? & Yes & 97.7\% & 99\% & 1.3\% (0.2\%, 2.4\%)* \\ 
  simple & Use full words vs. abbreviations? & Yes & 95.4\% & 96.2\% & 0.7\% (-0.9\%, 2.4\%) \\
  \hline
  complex & Clearly shows relationships? & Yes & 72.5\% & 83.7\% & 11.2\% (7.8\%, 14.5\%)* \\ 
  complex & Is the plot visually pleasing? & Somewhat & 30.5\% & 31\% & 0.5\% (-3.2\%, 4.3\%) \\ 
  complex & Is the plot visually pleasing? & Yes & 59.7\% & 60.5\% & 0.7\% (-3.2\%, 4.6\%) \\ 
  complex & Understandable without caption? & Yes & 76.3\% & 81.4\% & 5.1\% (1.8\%, 8.4\%)* \\ 
  complex & Legends and labels sufficient? & Yes & 77.6\% & 82.4\% & 4.8\% (1.6\%, 8\%)* \\ 
  complex & Text and labels large enough? & Yes & 90.5\% & 86.3\% & -4.2\% (-6.8\%, -1.6\%)* \\ 
  complex & Use full words vs. abbreviations? & Yes & 83.8\% & 85.5\% & 1.7\% (-1.2\%, 4.6\%) \\ 
   \bottomrule
\end{tabular}
\begin{tablenotes}
      \scriptsize
      \item Note: For each rubric item and response, the percentage of reviews indicating that response are shown. The last column gives the difference between the ggplot2 and base R arms and the 95\% confidence interval for that difference.
    \end{tablenotes}
\end{threeparttable}
\end{table}

\subsection{Types of complex plots made}

\begin{figure}
  \centering
  \captionsetup{width=.9\linewidth}
  \includegraphics[width=\textwidth]{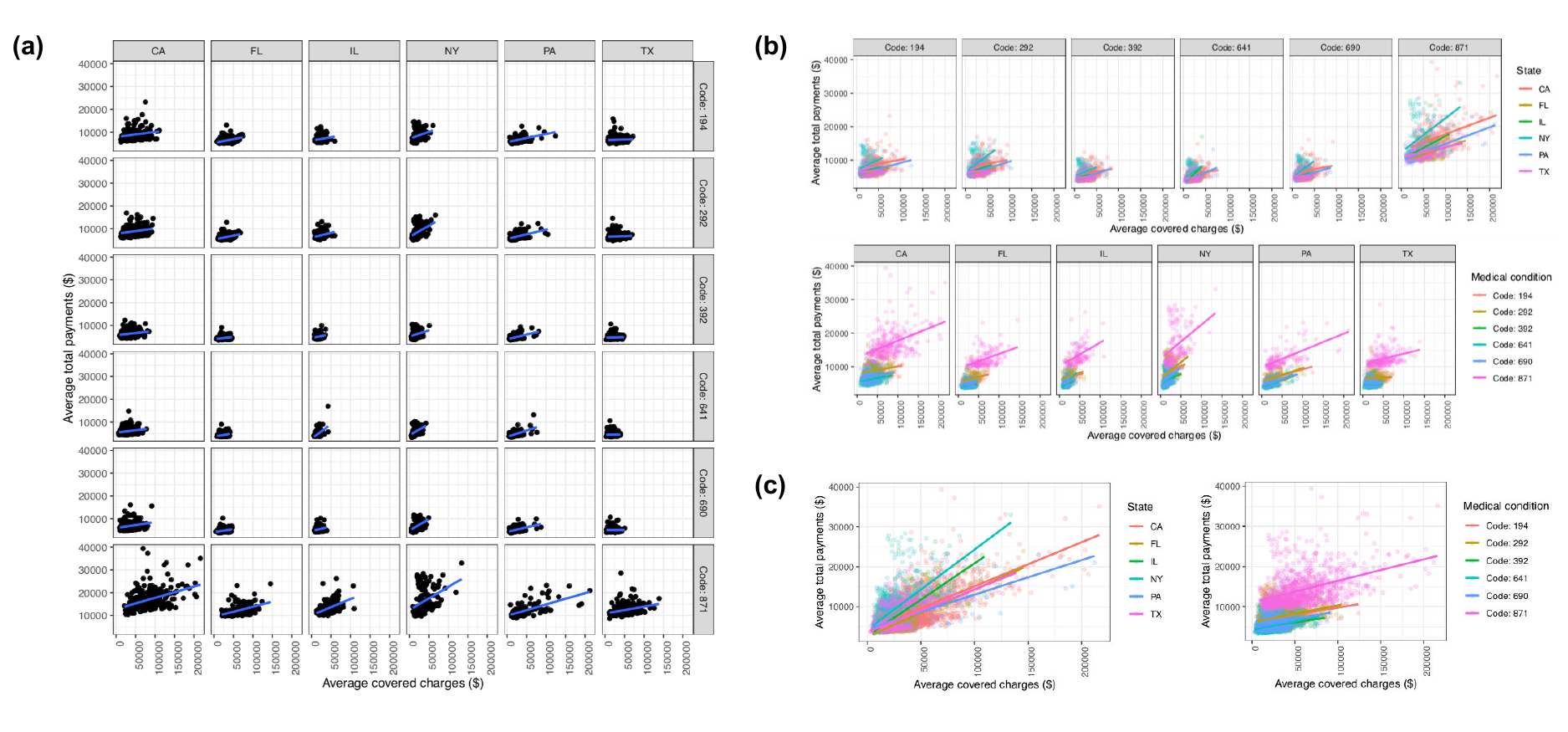}
  \caption{\textbf{Examples of types of complex plots.} (a) A typical 36 panel plot. (b) A typical 12 panel plot with coloring has panels for the 6 states colored by medical condition and panels for the 6 medical conditions colored by state. A typical 6 panel plot would show one of these two rows. (c) A typical 2 panel plot has one panel colored by the 6 states and a second panel colored by the 6 medical conditions.}
  \label{fig:complex_plot_types}
\end{figure}

Through our manual annotation of the complex plots, we were able to categorize the different types of student plots. Examples of the different types of plots are shown in Figure~\ref{fig:complex_plot_types}. The prevalence of these plot types in the base R and ggplot2 arms are shown in Table~\ref{tab:complex_plot_types}. In these results, we count base R and ggplot2 plots according to our manual annotation of the plotting system used, not by the actual experimental arm in which the student was enrolled.

Before completing the annotations, we hypothesized that the percentage of students making the full 6-by-6 panel of 36 scatterplots would be much higher in the ggplot2 arm because of the ease of syntax within the \texttt{facet\_grid()} function used for creating panels by categorical variables. This was indeed the case as 54.3\% of the ggplot2 submissions were 36 panel plots, compared to 31.9\% for base R plots (Table~\ref{tab:complex_plot_types}). The 6-by-6 panel of scatterplots is of pedagogical interest because this figure allows students to fully explore the interaction between the two categorical variables (medical condition and state). Although a 36 panel plot is the most concise formulation in the ggplot2 system, a 12 or 6 panel scatterplot that colors points by the remaining categorical variable (the one not used to define the panel) is perhaps more effective for making visual comparisons (Figure~\ref{fig:complex_plot_types}b). Such a figure places trends to be compared on the same plot, which facilitates comparisons more than the 36 panel plot. We see that all plot types aside from the 36 panel plot were more likely to be made in the base R submissions (Table~\ref{tab:complex_plot_types}). This may suggest interesting differences between the systems in how students process or approach the syntax needed to create such figures. We also see that the 36 panel, the colored 12 panel, and colored 6 panel plots were the most likely to be rated as clearly showing the intended relationship, as measured by the overall clarity rubric item (Table~\ref{tab:clarity_complex_plot_types}). Ratings of clarity in these plot type subgroups are uniformly higher for plots made in ggplot2.

\begin{table}[ht]
\scriptsize
\centering
\caption{\textbf{Types of complex plots.} For the 6 most common plot types, the percentage of submissions in which that type of plot was made is shown. The last column gives the difference between the ggplot2 and base R arms and the 95\% confidence interval for that difference.}
\label{tab:complex_plot_types}
\begin{tabular}{lrrl}
  \toprule
Plot type & Base R & ggplot2 & ggplot2 - base R \\ 
  \hline
36 panels & 31.9\% & 54.3\% & 22.3\% (16.1\%, 28.5\%)* \\ 
  12 panels, no color & 8.2\% & 3.2\% & -5.1\% (-8.3\%, -1.8\%)* \\ 
  12 panels, color & 1.9\% & 0.1\% & -1.7\% (-3.3\%, -0.1\%)* \\ 
  6 panels, no color & 4\% & 2\% & -2\% (-4.4\%, 0.5\%) \\ 
  6 panels, color & 42\% & 36.7\% & -5.4\% (-11.7\%, 1\%) \\ 
  2 panels, color & 5.6\% & 1.7\% & -3.9\% (-6.6\%, -1.1\%)* \\ 
   \bottomrule
\end{tabular}

\end{table}

\begin{table}[ht]
\scriptsize
\centering
\caption{\textbf{Clarity of the different types of complex plots.} For the 6 most common plot types, we show the percentage of submissions of each type that were judged to clearly show the intended relationship. Numbers in parentheses indicate sample sizes.}
\label{tab:clarity_complex_plot_types}
\begin{tabular}{lll}
  \toprule
Plot type & Base R & ggplot2 \\ 
  \hline
36 panels & 81.3\% (279/343) & 88.9\% (735/827) \\ 
  12 panels, no color & 64.4\% (56/87) & 73.5\% (25/34) \\ 
  12 panels, color & 81.8\% (9/11) &  \\ 
  6 panels, no color & 46.2\% (18/39) & 60\% (15/25) \\ 
  6 panels, color & 79.3\% (391/493) & 78.7\% (470/597) \\ 
  2 panels, color & 45.6\% (31/68) & 56.4\% (22/39) \\ 
   \bottomrule
\end{tabular}
\end{table}

\section{Discussion}

We performed a randomized trial in a group of beginner learners to understand their perceptions of statistical graphics made in the base R and ggplot2 systems. We find that for displaying bivariate relationships at an aggregate level and across strata, students using the ggplot2 system create graphics with slightly higher aesthetic appeal and greater scientific clarity, where clarity is measured by the questions ``Does the plot clearly show the relationship?'', ``Can the plot be understood without a figure caption?'', and ``Are the legends and labels sufficient to explain what the plot is showing?''. The clarity increase is greater when students attempted the more complex task of trying to depict a bivariate relationship across strata. We also observed that students were more likely to create plots in the assigned system when using ggplot2, suggesting a preference for ggplot2 due to factors we have not measured in our experiment. 

We also find that students are more likely to explore more complex interactions between variables when using ggplot2 than base R. Specifically, we saw a higher rate of students creating a full grid of scatterplots to answer the complex question when using ggplot2 than when using base R. This is in line with the relatively straightforward syntax for creating faceted plots within the ggplot2 framework. For the same figure to be made in base R, the students would have to used two nested for-loops, which may be an idea with which they are less comfortable. Despite the increased programming skill required, the most common type of plot made in the base R arm was a 6 panel figure that did require the use of a single for-loop.


Our results indicate that ggplot2 may slightly outperform base R, particularly as students move to faceted plots across multiple conditions. This provides evidence in favor of those advocating for the use of ggplot2 in introductory classes. The relatively small differences also suggest that both plotting systems can be capably used by beginning users to display scientific information. 

The scope of our plotting assignment is limited in terms of the breadth of statistical graphics that are used in practice, but it does cover the concepts of bivariate relationships and stratification, which are core ideas in data analysis in introductory statistics courses. The observed increase in reported scientific clarity for ggplot2 figures suggests that students have more favorable evaluations of these plots than plots made in base R. It is unclear whether this perceived increase in clarity is actually a result of more favorable aesthetic evaluations, but even if this is the case, students may be able to extract more scientific meaning from these plots simply because they are more comfortable with this plotting style. 

The strongest effects we observed in the data was in the type of plots made for faceted analysis. Students were significantly more likely to make the full 36 panel faceted figure in the complex case using the ggplot2 system.  While it does allow students to fully explore the interaction between the two categorical variables, it is not necessarily the most effective visualization because it requires the viewer to jump their eyes back and forth between panels to compare trends. A more effective plot collapses some of the panels by adding color, which was a more frequently made plot in the base R arm. These observations suggest student ease with the syntax used to create scatterplot grids in ggplot2. However, this ease may come at a price in terms of encouraging conscious efforts at the most effective visualizations. Although students seem to favor ggplot2 in terms of clarity and aesthetics, educators should be careful to continually emphasize the principles behind effective data visualization for communicating informative results.


\section{Other declarations}

\textbf{Data and code availability:} Code and data to reproduce the analyses here is available at \url{https://github.com/lmyint/ggplot_base}.

\noindent
\textbf{Funding:} This work was supported by National Institutes of Health grant R01GM115440.

\noindent
\textbf{Ethics:} We received approval to analyze this data from the Johns Hopkins Bloomberg School of Public Health: IRB number 00005988.

\newpage

\bibliographystyle{apalike.bst}
\bibliography{references}

\end{document}